\documentclass{JHEP3}
\usepackage{graphicx,amsmath,amssymb}
\usepackage{epsfig,multicol}
\usepackage{epsf}
\usepackage{epstopdf}
\input{epsf.sty}

\usepackage{graphicx}
\usepackage{amssymb}
\usepackage{amsmath}

\def\p{\partial}

\def\half{{1\over 2}}
\def\({\left(}
\def\){\right)}
\def\[{\left[}
\def\]{\right]}

\def\k{{\bf k}}
\def\x{{\bf x}}

\def\e{\begin{equation}}
\def\q{\end{equation}}
\def\m{\begin{eqnarray}}
\def\n{\end{eqnarray}}

\title{The trouble with asymptotically safe inflation}
\author{Chao Fang,\ and Qing-Guo Huang \footnote{huangqg@itp.ac.cn}
\\\small{\em
Kavli Institute for Theoretical Physics China (KITPC),\\ State Key Laboratory of Theoretical Physics,\\ Institute of Theoretical Physics, Chinese Academy of Science,\\ Beijing 100190, People's Republic of China} }

\abstract{
In this paper we investigate the perturbation theory of the asymptotically safe inflation and we find that all modes of gravitational waves perturbation become ghosts in order to achieve a large enough number of e-folds. Formally we can calculate the power spectrum of gravitational waves perturbation, but we find that it is negative. It indicates that there is serious trouble with the asymptotically safe inflation.
}


\begin{document}

\section{Introduction}

As a promising quantum theory of gravity, asymptotic safety of gravity was suggested in \cite{Weinberg:1977,Weinberg:1979}, and a version of perturbation theory was presented in \cite{Niedermaier:2006wt}. Such a theory is asymptotically safe, which implies that it might be ultraviolet (UV) complete. Recently ones showed that in the asymptotically safe gravity the renormalization group (RG) flows has a fixed point with three-dimensional UV critical surface of trajectories attracted to the fixed point in the UV limit, or equivalently at short distances \cite{Codello:2007bd,Codello:2008vh,Benedetti:2009rx,Benedetti:2009gn}. 

Since the asymptotically safe gravity has a brilliant behavior at short distances, it can be naturally applied to explore the physics in the early universe \cite{Bonanno:2001xi,Reuter:2005kb}, in particular inflation \cite{Guth:1980zm}. Recently an interesting version of asymptotically safe inflation was proposed by Weinberg \cite{Weinberg:2009wa} where a de Sitter solution is allowed and the instabilities are also naturally introduced to terminate inflation once including time dependence in the Hubble parameter. It is an attractive model in which any inflaton fields are not needed. However, assuming inflation happened at the UV fixed point, Weinberg found that inflation cannot last enough number of e-folds in some well-known examples with asymptotic safety. In \cite{Tye:2010an} Tye and Xu proposed that the asymptotically safe gravity should also reproduce present Newton's coupling constant and the cosmological constant driving current cosmic acceleration. In this scenario, inflation scale should be away from fixed point and enough e-folding number for inflation can be achieved with some fine-tuning. 

As we known, inflation not only is an elegant paradigm for solving several puzzles in hot big bang model, but also provides a natural explanation to the anisotropies in cosmic microwave background radiation and formation of large-scale structure in our universe. It is quite interesting for us to investigate the perturbation theory of the asymptotically safe inflation in this paper. We find that the asymptotically safe inflation suffers from a serious problem: the ghosts will make this model unacceptable. 

Our paper is organized as follows. In Sec.~2 the asymptotically safe inflation will be briefly reviewed. The  gravitational waves perturbation will be calculated in Sec.~3. Discussion is contained in Sec.~4. The action for the scalar curvature perturbation is given in the Appendix.

\section{The dynamics of asymptotically safe inflation}

Let's consider the following action 
\m
S=-\int d^4 x \sqrt{-g} \[{\mu^2\over g_N}(2\lambda \mu^2-R)+{1\over 2s}C^2-{\omega\over 3s}R^2\],
\label{action}
\n
here $\mu$ is the energy scale, $g_N$, $\lambda$, $\omega$ and $s$ are dimensionless parameters. $R$ is the Ricci scalar, $C$ is Weyl tensor and $C^2=R^{\mu\nu\rho\sigma}R_{\mu\nu\rho\sigma}-2R^{\mu\nu}R_{\mu\nu}+R^2/3$. $s$ is assumed to be positive for damping the Euclidean functional integral.

Following \cite{Weinberg:2009wa,Tye:2010an}, the dynamics of inflation govern by the action (\ref{action}) is 
\m
-{\lambda \over g_N}+{3\over g_N} {H^2\over \mu^2}+{6\omega\over s}{6H^2\dot H-\dot H^2+2H\ddot H\over \mu^4}=0.
\label{eom}
\n
If $\omega=0$, we can get a de Sitter solution. However, $R^2$ term may introduce an instability around the de Sitter solution. Similar to \cite{Weinberg:2009wa}, an ansatz for $H$ around de Sitter solution is proposed to be  
\m
H(t)=\bar H+\delta H(t),
\label{ht}
\n
with $\delta H(t)=\Lambda e^{\xi {\bar H}t}$ and
\m
{\bar H\over \mu}=\sqrt{\lambda\over 3}.
\n
Similar to slow-roll inflation, we also introduce the slow-roll parameters as follows 
\m
\epsilon=-{\dot H\over H^2},\quad \eta={\dot \epsilon\over \epsilon H}.
\n
From Eq.~(\ref{ht}),
\m
\dot H=\xi\bar H \delta H,\quad \ddot H=\xi^2 {\bar H}^2 \delta H, 
\n
and then
\m
\epsilon&=& -{\dot H\over H^2}\simeq -\xi {\delta H\over \bar H},\label{epsilon} \\
\eta&=&{\dot \epsilon\over \epsilon H}=2\epsilon+{\ddot H\over H\dot H}\simeq 2\epsilon+\xi.
\n
From Eq.~(\ref{epsilon}), we find $\delta H=-\epsilon{\bar H}/\xi$, and then 
\m
H=(1-{\epsilon\over \xi}){\bar H}. 
\label{hhb}
\n
Requiring $|\delta H/{\bar H}|\ll 1$ yields $\epsilon\ll \xi\lesssim 1/60$, and therefore $\eta\simeq \xi$.

On the other hand, from Eqs.~(\ref{eom}) and (\ref{ht}), we obtain 
\m
\xi^2+3\xi+{\kappa\over 2\omega}=0,
\label{bxi}
\n
where 
\m
\kappa={3s\over \lambda g_N}. 
\n
If all roots of Eq.~(\ref{bxi}) have Re$(\xi)<0$, the inflation govern by the vacuum energy $2\lambda\mu^4/g_N$ will last forever. This is not the case we are interested in. We will focus on the case with root of Eq.~(\ref{bxi}) satisfying Re$(\xi)>0$. It represents instabilities of the system, and the inflation will only last for $1/\hbox{Re}(\xi)$ e-folds. Eq.~(\ref{bxi}) has two solutions and one of them satisfies ${\rm Re}(\xi)>0$ if $\omega<0$, namely
\m
\xi={\sqrt{9-2\kappa/\omega}-3\over 2}.
\n
If $-{\kappa/ (2\omega)}\ll 1$, we obtain  
\m
\xi\simeq -{\kappa\over 6\omega}, 
\n
and then the total number of e-folds becomes 
\m
N_t\simeq {1\over \xi}\simeq -{6\omega\over \kappa}.  
\label{nt}
\n 
For getting enough e-folding number, namely $N_t\gtrsim 60$, we have $\kappa\lesssim -\omega/10$ which indicates that $\kappa$ is one order of magnitude smaller than $-\omega$ at least.

In \cite{Niedermaier:2006wt} the $\beta$ functions for $s$ and $\omega$ are given by 
\m
{ds\over d\ln \mu}&=&-{13.3\over (4\pi)^2}s^2,\\
{d\omega\over d\ln \mu}&=&-{s\over (4\pi)^2}{25+1098\omega+200\omega^2\over 60}, 
\n
which indicate that $s$ is asymptotically free and $\omega$ has a stable fixed point at 
\m
\omega_*=-0.0228. 
\label{omf}
\n
At the fixed point with $\omega=\omega_*$ and $s=0$, the two-dimensional RG flow of $g_N$ and $\lambda$ becomes \cite{Niedermaier:2009zz}
\m
\mu{d g_N\over d\mu}&=&2g_N-\gamma_1 g_N^2+{\cal O}(g_N^3), \\
\mu{d\lambda\over d\mu}&=&-2\lambda+a_1 g_N+a_2 g_N\lambda+a_3g_N^2+{\cal O}(g_N^2\lambda), 
\n
where 
\m
a_1={2u_1^*\over (4\pi)^2},\quad \gamma_1=-a_2={2u_2^*\over (4\pi)^2}, 
\n
and $u_1^*=1.38$, $u_2^*=0.73$. The solution of the above renormalization group flow is given by \cite{Tye:2010an}
\m
{g_N\over (4\pi)^2}&=&{(\mu/\mu_0)^2\over 1+u_2^*(\mu/\mu_0)^2}, \\
\lambda&=&{(\mu/\mu_\lambda)^{-4}+u_1^*\over 2(\mu/\mu_0)^{-2}+2u_2^*}, 
\label{lambda} 
\n
where $\mu_0$ and $\mu_\lambda$ are free parameters. In the infrared (IR) limit $(\mu\rightarrow 0)$, the Einstein Hilbert term in the action (\ref{action}) becomes 
\m
{\mu_\lambda^4-\mu_0^2 R\over (4\pi)^2}.
\n 
In \cite{Tye:2010an}, both the present Newton's coupling and cosmological constant were proposed to be recovered in the IR limit of the asymptotically safe gravity.  
Comparing to the standard form $\Lambda-{M_p^2\over 2}R$, where $M_p$ is the reduced Planck scale and $\Lambda=7.45\times 10^{-121}M_p^4$ is today's cosmological constant, we can fix $\mu_\lambda$ and $\mu_0$ to be 
\m
\mu_\lambda\simeq 3\times 10^{-30}M_p,
\n 
and 
\m
\mu_0=\sqrt{8}\pi M_p.
\label{mu0}
\n 
In the UV limit $(\mu\gg \mu_0\sim M_p)$, $g_N$ and $\lambda$ flow to the fixed point with $g_N^*/(4\pi)^2=1/u_2^*=1.37$ and $\lambda^*=u_1^*/(2u_2^*)=0.95$. Since UV limit corresponds to a very high energy scale, in \cite{Tye:2010an} inflation is assumed to be away from fixed point, on scales $H_0\ll \mu\lesssim M_p$, where $H_0$ is the today's Hubble constant. From Eq.~(\ref{lambda}), we obtain 
\m 
\lambda\simeq {u_1^*\over 2}\({\mu\over \mu_0}\)^2.
\n 
Considering $H/\mu\simeq \sqrt{\lambda/3}$, we get 
\m
H\simeq \sqrt{u_1^*\over 6}{\mu^2\over \mu_0}, 
\label{muh}
\n 
and thus 
\m
{g_N\over (4\pi)^2}&\simeq& \sqrt{3\over 4u_1^* \pi^2} {H\over M_p}, \label{gnh} \\
\lambda&\simeq& \sqrt{3 u_1^*\over 16 \pi^2}{H\over M_p}.
\n
Therefore we find 
\m
\kappa\simeq 0.52 \({H\over M_p}\)^{-2}s,
\n 
and $N_t\gtrsim 60$ requires 
\m
s\lesssim -{\omega\over 5.2} \({H/ M_p}\)^2.
\n 
For $\omega=\omega_*$, $s\lesssim 4\times 10^{-3} \({H/ M_p}\)^2$. Considering that inflation scale should be much lower than the Planck scale, $s$ should be very small and a fine-tuning is called for.

\section{Gravitational waves perturbation in asymptotically safe inflation}
\label{sgw}

In this section, we focus on the gravitational waves perturbation. The perturbed metric is given by 
\m 
ds^2=a^2(\tau)\[-d\tau^2+(e^h)_{ij}dx^idx^j \], 
\n
where $\tau=\int dt/a$ is the conformal time.
Up to quadratic order, we have 
\m 
(e_h)_{ij}\equiv\delta_{ij}+h_{ij}+{1\over2}h_{ik}h^k_j, 
\n
and 
\m 
(e^h)^{ij}\equiv\delta^{ij}-h^{ij}+{1\over2}h^{ik}h^j_k. 
\n
Here $(e_h)_{ij}$ is just an abstract notation. This approach is equal to the the one simply writing the
tensor perturbation as $(\delta_{ij}+h_{ij})$ in the linear order. The symmetric tensor $h_{ij}$ has six degrees of freedom, but the gravitational waves perturbation is traceless, 
\m
\delta^{ij}h_{ij}=0, 
\n 
and transverse 
\m
\p^i h_{ij}=0. 
\n 
With these four constraints, there remains two physical degrees of freedom, or polarizations. One can easily find $\sqrt{-g} =a^4$ and then the dynamics of gravitational waves perturbation is governed by the action 
\m
S=\int d\tau dx^3 \left\{ \({a^2\mu^2\over 4g_N}+(2-\epsilon){\omega\over s}{\cal H}^2\)\[h_{ij}'^2-(\p_l h_{ij})^2\]- {1\over 4s} \(h_{ij}''-\Delta h_{ij}\)^2 \right\}, 
\label{actiongw}
\n
where $\Delta\equiv \p_i\p_i$, the primes denote the derivatives with respect to the conformal time $\tau$ and ${\cal H}\equiv a'/a$. Combining Eqs.~(\ref{mu0}), (\ref{muh}) and (\ref{gnh}), we find ${\mu^2\over 4g_N}={M_p^2\over 16}$. One can easily see that the terms with ${a^2\mu^2\over 4g_N}$ and ${\omega\over s}$ in the above action come from the Einstein Hilbert term and $R^2$ terms respectively, and the term with ${1\over s}$ comes from the term of $C^2$. 
From the action (\ref{actiongw}), the equation of motion for $h_{ij}$ takes the form 
\m 
&&{a^2\mu^2\over g_N}[h_{ij}''+2{\cal H}h_{ij}'-\Delta h_{ij}]+4(2-\epsilon){\omega\over s}{\cal H}^2 \[ h_{ij}''+2{\cal H}\(1-\epsilon-{\epsilon\eta\over 2(2-\epsilon)}\)h_{ij}'-\Delta h_{ij}\] \nonumber \\ 
&&+{1\over s}\(\p_\tau^2-\Delta\)^2 h_{ij}=0. 
\label{gw}
\n

Using Eq.~(\ref{hhb}), we obtain 
\m
{s\mu^2 a^2\over g_N {\cal H}^2}&\simeq& \kappa(1+2\epsilon/\xi), \\
{\lambda \mu^2 a^2\over {\cal H}^2}&\simeq& 3(1+2\epsilon/\xi). 
\n
Since $\epsilon\ll \xi\lesssim 1/60$, the terms with $\epsilon$ can be ignored and then the action in Eq.~(\ref{actiongw}) becomes 
\m
S=\int d\tau dx^3 \left\{ \({\kappa\over 4}+2\omega\) {{\cal H}^2\over s}\[h_{ij}'^2-(\p_l h_{ij})^2\]- {1\over 4s} \(h_{ij}''-\Delta h_{ij}\)^2 \right\}.  
\label{actsgw}
\n
Considering $10\kappa\lesssim -\omega\ll 1$, the term of $\kappa$ becomes subdominant and can be neglected in the case with enough number of e-folds, and then the equation of motion for $h_{ij}$ is simplified to 
\m 
8\omega {\cal H}^2 \[ h_{ij}''+2{\cal H}h_{ij}'-\Delta h_{ij}\] + h_{ij}''''-2\Delta h_{ij}''+\Delta^2h_{ij}=0. 
\label{gws}
\n
Since there are fourth derivative with respect to the time $\tau$, one more degree of freedom is expected to appear. In order to make it more clear, we introduce a Lagrange multiplier $\lambda_{ij}$ and consider the following equivalent action 
\m
S&=&\int d\tau dx^3 \left\{ 2{\omega\over s}{\cal H}^2\[h_{ij}'^2-(\p_l h_{ij})^2\] \right. \nonumber \\
&-& \left. {1\over 4s} \({Q_{ij}'}^2-2\p_l h_{ij}' \p_l h_{ij}'+(\Delta h_{ij})^2\) - \lambda_{ij}(Q_{ij}-h_{ij}')\right\}.  
\label{actiongwm}
\n 
Here $h_{ij}$ and $Q_{ij}$ denote the physical degrees of freedom and their conjugate momenta take the form 
\m
\pi^h_{ij}&=&{4\omega \over s}{\cal H}^2 h_{ij}' -{1\over s}\Delta h_{ij}'+\lambda_{ij}, \\
\pi^Q_{ij}&=&-{1\over 2s}Q_{ij}'. 
\n
Considering $\omega<0$ and $s>0$, both $h_{ij}$ and $Q_{ij}$ are ghost fields. From the above action, the equations of motion for $\lambda_{ij}$ and $Q_{ij}$ are given by 
\m
Q_{ij}&=&h_{ij}', \\
\lambda_{ij}&=&{1\over 2s}Q_{ij}'', 
\n
and then 
\m
\pi^h_{ij}&=&{4\omega \over s}{\cal H}^2 h_{ij}' -{1\over s}\Delta h_{ij}'+{1\over 2s}h_{ij}''', \\
\pi^Q_{ij}&=&-{1\over 2s}h_{ij}''. 
\n
Formally we can quantize the field $h_{ij}(\tau,\x)$ as follows 
\m
h_{ij}(\tau,\x)=\int {d^3k\over (2\pi)^{3/2}} \[\sum_{\lambda=\pm 2} \({\hat a}_\k^\lambda h_k^{(1)}(\tau)+{\hat b}_\k^\lambda h_k^{(2)}(\tau)\) e_{ij}({\bf k},\lambda)e^{i {\bf k}\cdot {\bf x}}+{\rm h.c.}\], 
\n
where ${\hat a}$ and ${\hat b}$ are two operators which satisfy  
\m
[{\hat a}_{\k_1}^{\lambda_1},{\hat a}_{\k_2}^{\lambda_2 \dagger}]=-\delta_{\lambda_1 \lambda_2}\delta^{(3)}(\k_1-\k_2), \quad [{\hat b}_{\k_1}^{\lambda_1},{\hat b}_{\k_2}^{\lambda_2 \dagger}]=-\delta_{\lambda_1 \lambda_2}\delta^{(3)}(\k_1-\k_2), 
\n
all other commutators vanish, and $e_{ij}({\bf k},\lambda=\pm 2)$ are two physical time independent polarization tensors  
\m
e_{ij}=e_{ji},\quad e_{ii}=0,\quad k_i e_{ij}=0. 
\n
For simplicity, we normalize the tensor $e_{ij}$ to be  
\m
\sum_{\lambda=\pm 2} e_{ij}^*({\bf k},\lambda) e_{ij}({\bf k},\lambda)=2. 
\n
Canonical quantization imposes the the following commutation relations 
\m
[h_{ij}(\tau,\x_1),\pi^h_{ij}(\tau,\x_2)]=2i\delta^{(3)}(\x_1-\x_2), 
\n
and 
\m
[Q_{ij}(\tau,\x_1),\pi^Q_{ij}(\tau,\x_2)]=2i\delta^{(3)}(\x_1-\x_2), 
\n
all other commutators being zero and the factor 2 coming from the fact that both $h_{ij}$ and $Q_{ij}$ represent two degrees of freedom. These two commutation relations imply 
\m
h_k^{(1)}\[(1+{k^2\over 4\omega {\cal H}^2})h_k'^{(1)*}+{1\over 8\omega {\cal H}^2}h_k'''^{(1)*}\]&+&h_k^{(2)}\[(1+{k^2\over 4\omega {\cal H}^2})h_k'^{(2)*}+{1\over 8\omega {\cal H}^2}h_k'''^{(2)*}\] \nonumber \\ 
-{\rm c.c.}&=&-i{s \over 4\omega {\cal H}^2}, \label{nh}\\
h_k'^{(1)} h_k''^{(1)*}+h_k'^{(2)} h_k''^{(2)*}-{\rm c.c.}&=&2si. \label{nq}
\n
We need to remind that both mode functions $h_k^{(1)}$ and $h_k^{(2)}$ and their complex conjugates are four independent solutions of the equation of motion for $h_{ij}$ in Eq.~(\ref{gws}) which reads 
\m 
8\omega {\cal H}^2 \[ h_{ij}''+2{\cal H}h_{ij}'+k^2 h_{ij}\] + h_{ij}''''+2k^2 h_{ij} h_{ij}''+k^4 h_{ij}=0. 
\n
Similar to \cite{Clunan:2009er,Deruelle:2012xv}, the above equation can be factorized as follows 
\m
{d^2\over dz^2}v_k^{(1)}+\(1-{2\over z^2}\)v_k^{(1)}&=&0, \label{v1} \\
{d^2\over dz^2}v_k^{(2)}+\(1+{8\omega \over z^2}\)v_k^{(2)}&=&0, \label{v2}
\n
where $z=-k\tau$ and $v_k^{(i)}=h_k^{(i)}/z$ for $i=1,2$. We choose the positive frequency modes which define a ``Bunch-Davies" vacuum state $|0\rangle$: ${\hat a}_k|0\rangle={\hat b}_k|0\rangle=0$. In the sub-horizon limit ($z\rightarrow \infty$), the equations of motion for $v_k^{(i)}$ are reduced to be those in the Minkowski space-time and thus  
\m
v_k^{(1)}\sim v_k^{(2)}\sim e^{iz}. 
\n
Therefore $h_k^{(i)}$ become 
\m
h_k^{(1)}\sim h_k^{(2)}\sim ze^{iz}. 
\n
On the other hand, using Eqs.~(\ref{v1}) and (\ref{v2}), the canonical quantization conditions in Eqs.~(\ref{nh}) and (\ref{nq}) become 
\m
h^{(1)}_{k}{dh^{(1)*}_{k}\over dz}-c.c=-h^{(2)}_{k}{dh^{(2)*}_{k}\over dz}-c.c={i2sz^2\over(2+8\omega)k^3}. 
\n
Therefore, in the sub-horizon limit, we have 
\m
h_k^{(1)}&=&i \sqrt{s\over (2+8\omega)k^3}ze^{iz}, \\
h_k^{(2)}&=&\sqrt{s\over (2+8\omega)k^3}ze^{iz}. 
\n
Up to a phase, the solution of $h_k^{(i)}$ can be written down by 
\m 
h^{(1)}_{k}&=&\sqrt{{s\over k^3(2+8\omega)}}(1+iz)e^{iz}, \\
h^{(2)}_{k}&=&\sqrt{{s\over k^3(2+8\omega)}}\sqrt{\pi\over 2}e^{i\pi\nu\over 2}z^{{3\over2}}H^{(1)}_\nu(z)
\n
where $\nu\equiv {1\over2}\sqrt{1-32\omega}$ and $H_\nu^{(1)}$ is the Hankel function of the first kind.

Now we can calculate the power spectrum $P_T(k,z)$ of the gravitational waves perturbation which is defined by 
\m 
\langle0| h_{ij}(\tau;\x_1) h_{ij}(\tau;\x_2)|0\rangle=\int {d^3k\over 4\pi k^3} P_T(k;z) e^{i\k\cdot (\x_1-\x_2)}. 
\n
From the above definition, we have 
\m 
P_T(k;z)&=&-{k^3\over \pi^2}\( |h^{(1)}_k|^2+|h^{(2)}_k|^2\) \nonumber\\
&=&-{s\over 2\pi^2(1+4\omega)}\(1+z^2+{\pi\over2}z^3 |e^{i\pi\nu\over
2}H^{(1)}_\nu(z)|^2\), 
\n
and hence the power spectrum of the gravitational waves perturbation is 
\m 
P_T\equiv P_T(k;z\rightarrow 0)=-{s\over 2\pi^2(1+4\omega)}. 
\n
For $\omega=\omega_*=-0.0228$, $P_T\simeq -0.056 s$ which is negative! It indicates that there is a serious trouble with the asymptotically safe inflation with enough number of e-folds. More discussion on it is given in the next section.

\section{Discussion}

Asymptotically safe inflation is proposed to be a quite attractive inflation model in which the inflaton is not needed. We clearly point out that the instability can come from the term of $R^2$ as long as $\omega<0$ in (\ref{action}). In this case the asymptotically safe inflation can naturally end. On the other hand, in order to achieve a large number of e-folds, $\kappa$ has to be one order of magnitude smaller than $-\omega$ at least.

\begin{figure}[h]
\centering
\includegraphics[width=12cm]{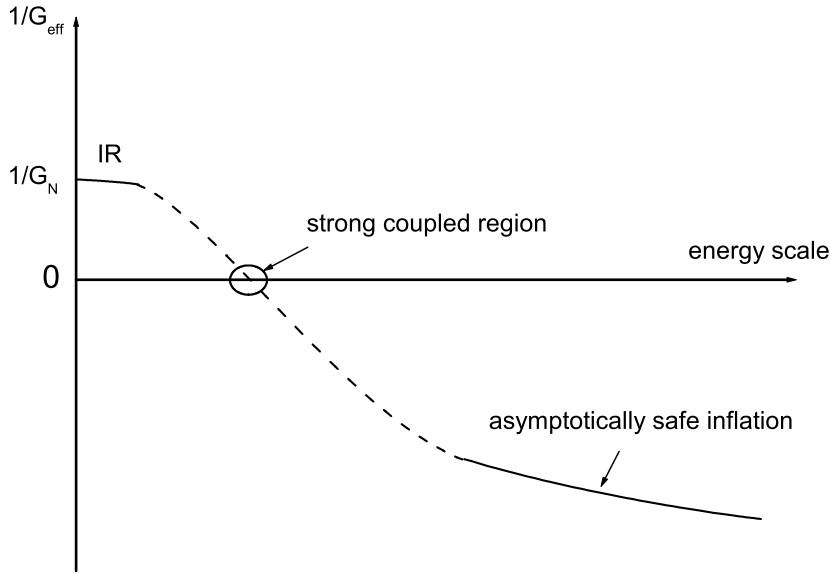}
\caption{\label{fig:safe} The flow of $G_{\rm eff}$.}
\end{figure}

We also study the gravitational waves perturbation in the asymptotically safe inflation in detail. Considering $\kappa\lesssim -\omega/10$, we find that the contribution to the action of $h_{ij}$ from the term of $R^2$ is much larger than that from the Einstein Hilbert term. Since $\omega<0$, all perturbation modes of gravitational waves become ghosts! It indicates that the asymptotically safe inflation breaks the unitarity. On the other hand, let's re-consider the action in (\ref{actsgw}) and the coefficient $\({\kappa\over 4}+2\omega\) {H^2\over s}$ can be interpreted as the effective coupling constant of graviton, namely 
\m
{1\over G_{\rm eff}}=32\pi \(\kappa+8\omega\){H^2\over s}. 
\n
In the limit of $\omega\rightarrow 0$, $G_{\rm eff}$ is nothing but the Newton's coupling constant $G_N=1/(8\pi M_p^2)$. During inflation, $\kappa\lesssim -\omega/10$ and then $G_{\rm eff}<0$. However, in the IR limit, Einstein gravity should be recovered and hence $G_{\rm eff}>0$. The flow of $G_{\rm eff}$ in the asymptotically safe inflation can be illustrated in Fig.~\ref{fig:safe}.
Between UV and IR, the effective coupling constant $G_{\rm eff}$ must go to infinity around some energy scale and the theory becomes strong coupled theory. It implies that such an asymptotically safe gravity cannot explain the physics from UV to IR. In a word, the maliciousness of ghosts emerging at the perturbed level breaks the beauty of asymptotically safe inflation and makes it unacceptable.

We also compute the action for the scalar curvature perturbation in the Longitudinal gauge. See the appendix \ref{ap}. We find that ghosts emerge as well. One possible extension to the asymptotically safe inflation in this paper is to consider the higher order curvature terms, such as $R^3$ and so on. It will be interesting to investigate the RG flow for such an extension. Another possibility is to involve a suitable scale field $\phi$ which can keep the asymptotic safety. These possible extensions will be left for the future.

\vspace{1.4cm}

\noindent {\bf Acknowledgments}

\vspace{.5cm}

QGH would like to thank Henry Tye for helpful discussions. 
This work is supported by the project of Knowledge Innovation
Program of Chinese Academy of Science and a grant from NSFC (grant NO. 10975167).


\appendix

\section{Appendix}
\label{ap}

We start with a general scalar perturbed metric about the flat FRW background as follows 
\m
ds^{2}=-\(1+\alpha\)dt^{2}-2a\(t\)\partial_{i}{\beta}dtdx^{i}+a^{2}\(t\)\(\delta_{ij}+2\zeta\delta_{ij}+2\partial_{i}\partial_{j}\gamma\). 
\n
The gauge transformations  are:
\m
\hat{\alpha}&=&\alpha-\dot{\delta t}, \\
\hat{\beta}&=&\beta-a^{-1}\delta t+a\dot{\delta x}, \\
\hat{\zeta}&=&\zeta-H\delta t, \\
\hat{\gamma}&=&\gamma-\delta x. 
\n
In this paper we will work on the Longitudinal gauge which corresponds to the gauge choice $\hat{\beta}=0$ and $\hat{\gamma}=0$ by setting the $\delta t=a\(\beta+a\dot{\gamma}\)$ and $\delta x=\gamma$. In this gauge, the line element becomes 
\m
ds^{2}=a^2(\tau)\[ -(1+2\alpha)d\tau^2+(1+2\zeta)\delta_{ij}dx^{i}dx^{j} \], 
\n
where we omitted the hat for perturbed quantities and $\tau$ is conformal time which is related to time $t$ by
\m
d\tau={dt\over a}. 
\n 
The equations of motion for $\alpha$ and $\zeta$ are governed by the action in Eq.~(\ref{action}). In the Longitudinal gauge, after a lengthy but straightforward calculation, we obtain the action for perturbations as follows
\m
S=S_1+S_2+S_3,
\n
where 
\m
S_1=&&-\half \int d\tau d^3 x {\mu^2 a^2\over g_N} \[2{\cal H}^2 \(-{\lambda\mu^2 a^2\over {\cal H}^2}+9\) \alpha^2+ 12{\cal H}^2 \({\lambda \mu^2 a^2\over {\cal H}^2}-3\) \alpha\zeta \right.  \nonumber \\
&&\left. +6{\cal H}^2 \({\lambda\mu^2 a^2\over {\cal H}^2}-3+2\epsilon\)\zeta^2 -24{\cal H} \alpha \zeta'+4\(3\zeta'^2+2\alpha\Delta\zeta+\zeta\Delta\zeta\)\], 
\n
\m
S_2=-\half \int d\tau d^3 x {4\over 3s}\[(\Delta\alpha)^2-2\Delta\alpha \Delta\zeta+(\Delta\zeta)^2\] ,
\n
and 
\m
S_3=&&-\half \int d\tau d^3 x {\omega\over s}\{ 12{\cal H}^4 \[ 5(-6\epsilon+3\epsilon^2-2\epsilon\eta)\alpha^2+6(6\epsilon-3\epsilon^2+2\epsilon\eta)\alpha\zeta \right. \nonumber \\
&&\left. +(18\epsilon-33\epsilon^2+12\epsilon^3+12\epsilon\eta-10\epsilon^2\eta+2\epsilon\eta^2)\zeta^2\]+144 {\cal H}^3(4\epsilon-2\epsilon^2+\epsilon\eta)\alpha\zeta' \nonumber \\
&&+{\cal H}^2\[ -24\alpha'^2-48(1-2\epsilon)\alpha'\zeta'-24(2+5\epsilon)\zeta'^2  \right. \nonumber \\
&&\left. -32(2-\epsilon)\alpha\Delta\alpha-32(2-\epsilon)\alpha\Delta\zeta+16(2-\epsilon)\zeta\Delta\zeta \] \nonumber \\
&&+16{\cal H}\(3\alpha'\zeta''-\alpha'\Delta\alpha-2\alpha'\Delta\zeta+3\zeta'\Delta\alpha+6\zeta'\Delta\zeta\) \nonumber \\
&&-24\zeta''^2+16\alpha''\Delta\zeta+32\zeta''\Delta\zeta-{8\over 3}(\Delta\alpha)^2-{32\over 3}\Delta\alpha\Delta\zeta-{32\over 3}(\Delta\zeta)^2 \}, 
\n
From the above action, the equations of motion for $\alpha$ and $\zeta$ become 
\m
&&{s\mu^2a^2\over g_N {\cal H}^2} \[{1\over 2}\({\lambda\mu^2 a^2\over {\cal H}^2}-9\)\alpha-{3\over2}\({\lambda\mu^2 a^2\over {\cal H}^2}-3\)\zeta+{3\over {\cal H}}\zeta'-{1\over {\cal H}^{2}}\Delta\zeta \]-{1\over 3{\cal H}^4}\Delta^{2}\alpha+{1\over 3{\cal H}^{4}}\Delta^{2}\zeta\nonumber\\
&&+\omega\[90\epsilon\alpha-54\epsilon\zeta-{12\over {\cal H}}\(1-\epsilon\)\alpha'-{12\over {\cal H}}(1+3\epsilon)\zeta' -{6\over {\cal H}^2}\alpha''+{6\over {\cal H}^{2}}\zeta''+{2\over {\cal H}^{2}}\(7-3\epsilon\)\Delta\alpha \right.\nonumber\\
&&\left.+{4\over {\cal H}^{2}}\Delta\zeta+{6\over {\cal H}^{3}}\zeta'''-{10\over {\cal H}^{3}}\Delta\zeta'-{2\over {\cal H}^{4}}\Delta\zeta''+{2\over 3{\cal H}^{4}}\Delta^{2}\alpha+{4\over 3{\cal H}^{4}}\Delta^{2}\zeta +{\cal O}(\epsilon^2,\epsilon\eta,\cdots) \]=0, \nonumber \\
\label{alpha}
\n
and 
\m
&&{s\mu^2a^2\over g_N {\cal H}^2}\[{3\over 2}\(-{\lambda\mu^2a^2\over {\cal H}^{2}}-3+2\epsilon\)\alpha+{3\over 2}\(-{\lambda\mu^{2}a^2\over {\cal H}^{2}}+3-2\epsilon\)\zeta-{3\over{\cal H}}\alpha'+{6\over{\cal H}}\zeta'+{3\over {\cal H}^{2}}\zeta'' \right. \nonumber\\
&&\left. -{1\over {\cal H}^{2}}\Delta\alpha-{1\over {\cal H}^{2}}\Delta\zeta \]+{1\over {3\cal H}^{4}}\Delta^{2}\alpha-{1\over 3{\cal H}^{4}}\Delta^{2}\zeta \nonumber\\
&&+\omega\[162\epsilon \alpha-54\epsilon\zeta-{12\over {\cal H}}(2-11\epsilon)\alpha' - {12\over {\cal H}}(2+3\epsilon)\zeta' -{6\over {\cal H}^{2}}\(3-4\epsilon\)\alpha''-{6\over {\cal H}^{2}}\(2+5\epsilon\)\zeta''\right.\nonumber\\
&&\left. +{2\over {\cal H}^{2}}\(7-5\epsilon\)\Delta\alpha+{4\over {\cal H}^{2}}\(1-2\epsilon\)\Delta\zeta+{10\over {\cal H}^{3}}\Delta\alpha'-{6\over {\cal H}^{3}}\alpha'''+{6\over {\cal H}^{4}}\zeta''''-{2\over {\cal H}^{4}}\Delta\alpha''-{8\over {\cal H}^{4}}\Delta\zeta'' \right.\nonumber\\
&&\left.+{4\over 3{\cal H}^{4}}\Delta^{2}\alpha+{8\over 3{\cal H}^{4}}\Delta^{2}\zeta+{\cal O}(\epsilon^2,\epsilon\eta,\cdots) \]=0. 
\label{zeta}
\n
Similar to Sec.~\ref{sgw}, ignoring the terms with $\epsilon$ and $\kappa$, the action for $\alpha$ and $\zeta$ becomes 
\m
S=&&-\half \int d\tau d^3 x \left\{ {\mu^2 a^2\over g_N} \[12{\cal H}^2 \alpha^2  -24{\cal H} \alpha \zeta'+4\(3\zeta'^2+2\alpha\Delta\zeta+\zeta\Delta\zeta\)\] \right. \nonumber \\
&&\left. +{4\over 3s}\[(\Delta\alpha)^2-2\Delta\alpha \Delta\zeta+(\Delta\zeta)^2\]  \right. \nonumber \\ 
&&\left.+{\omega\over s}[ {\cal H}^2\( -24\alpha'^2-48\alpha'\zeta'-48\zeta'^2-64\alpha\Delta\alpha-64\alpha\Delta\zeta+32\zeta\Delta\zeta \) \right. \nonumber \\
&&\left. +16{\cal H}\(3\alpha'\zeta''-\alpha'\Delta\alpha-2\alpha'\Delta\zeta+3\zeta'\Delta\alpha+6\zeta'\Delta\zeta\) \right. \nonumber \\
&&\left. -24\zeta''^2+16\alpha''\Delta\zeta+32\zeta''\Delta\zeta-{8\over 3}(\Delta\alpha)^2-{32\over 3}\Delta\alpha\Delta\zeta-{32\over 3}(\Delta\zeta)^2] \right\}, 
\n
We see that ghosts also emerge in the above action.

\end{document}